\documentclass[a4paper,12pt]{article}

\usepackage{amstext,amsmath,amssymb,amsthm}  

\usepackage{graphicx}  
\usepackage[comma,sort,authoryear]{natbib}
\usepackage{hypernat}  

\usepackage{vmargin}  
\usepackage{authblk}

\usepackage{xcolor}
\definecolor{dark-red}{rgb}{0.8,0.15,0.15}
\definecolor{dark-blue}{rgb}{0.15,0.15,0.6}
\definecolor{medium-blue}{rgb}{0,0,0.8}

\usepackage[pdftex,breaklinks,colorlinks]{hyperref}
\hypersetup{  
  pdftitle =
    {Shut up and let me think! Or why you should work on the foundations of quantum mechanics as much as you please},
  linkcolor={dark-red},
  citecolor={dark-blue},
  urlcolor={medium-blue}
}

\begin{document}

\numberwithin{equation}{section}
\numberwithin{figure}{section}
\allowdisplaybreaks[1]  

\title{Shut up and let me think! \\ Or why you should work on the foundations of quantum mechanics as much as you please}

\author[,1,2,3,4,5]{Pablo Echenique-Robba\footnote{{\footnotesize \
\href{mailto:pablo.echenique.robba@gmail.com}{\texttt{pablo.echenique.robba@gmail.com}} ---
\href{http://www.pabloecheniquerobba.com}{\texttt{http://www.pabloecheniquerobba.com}}}}}

\affil[1]{Instituto de Qu\'{\i}mica F\'{\i}sica Rocasolano, CSIC, Madrid, Spain}
\affil[2]{Instituto de Biocomputaci\'on y F{\'{\i}}sica de Sistemas Complejos (BIFI), Universidad de Zaragoza, Spain}
\affil[3]{Zaragoza Scientific Center for Advanced Modeling (ZCAM), Universidad de Zaragoza, Spain}
\affil[4]{Departamento de F{\'{\i}}sica Te\'orica, Universidad de Zaragoza, Spain}
\affil[5]{Unidad Asociada IQFR-BIFI, Madrid-Zaragoza, Spain}

\date{\today}

\maketitle

\begin{abstract}

If you have a restless intellect, it is very likely that you have played at
some point with the idea of investigating the meaning and conceptual
foundations of quantum mechanics. It is also probable (albeit not certain)
that your intentions have been stopped on their tracks by an encounter with
some version of the ``Shut up and calculate!'' command. You may have heard
that everything is already understood. That understanding is not your job. Or,
if it is, it is either impossible or very difficult. Maybe somebody explained
to you that physics is concerned with ``hows'' and not with ``whys''; that
whys are the business of ``philosophy'' ---you know, that dirty word. That
what you call ``understanding'' is just being Newtonian; which of course you
cannot ask quantum mechanics to be. Perhaps they also complemented these
useful advices with some norms: The important thing a theory must do is
predict; a theory must only talk about measurable quantities. It may also be
the case that you almost asked ``OK, and why is that?'', but you finally bit
your tongue. If you persisted in your intentions and the debate got a little
heated up, it is even possible that it was suggested that you suffered of some
type of moral or epistemic weakness that tend to disappear as you grow up.
Maybe you received some job advice such as ``Don't work in that if you ever
want to own a house''.

I have certainly met all these objections in my short career, and I think that
they are all just \emph{wrong}. In this somewhat personal document, I try to
defend that making sense of quantum mechanics is an exciting, challenging,
important and \emph{open} scientific endeavor. I do this by compulsively
quoting Feynman (and others), and I provide some arguments that you might want
to use the next time you confront the mentioned ``opinions''. By analogy with
the anti-rationalistic Copenhagen command, all the arguments are subsumed in a
standard answer to it: ``Shut up and let me think!''
\vspace{0.2cm}\\

\end{abstract}

\section{On a personal note}
\label{sec:personal}

\begin{flushright}
\parbox[h]{5in}{{\small
Science alone of all the subjects contains within itself the lesson of 
the danger of belief in the infallibility of the greatest teachers in the 
preceding generation\ldots Learn from science that you must doubt the experts. 
As a matter of fact, I can also define science another way: Science is the 
belief in the ignorance of experts.
\begin{flushright}
\vspace{-23.5pt}
--- \cite{Feynman1968}
\end{flushright}
}}
\end{flushright}

I have recently decided that I will spend some time and effort to try to come
up with a consistent and satisfying account of how QM (quantum mechanics)
works and what QM means. That is, consistent and satisfying to me.

My decision is based on a fact and a conviction: The first is that I enjoy the
topic immensely, the second is that I believe that a conceptual tidying up is
an urgent necessity in the field. Both circumstances combined make my decision
an easy bargain.

Although many scientists and philosophers certainly agree with me on this,
there is a strong and steady current of thought, and statements, that sees
such a tidying-up enterprise as futile and entirely irrelevant. Everything is
perfectly tidy for them. Their viewpoint (slightly caricaturized or not so
much, depending on who we are talking about) is that non-relativistic QM has
been completely understood an unspecified number of decades ago, that the
so-called ``measurement problem'' is no problem at all, and that all the
discussions about the different ``interpretations'' of QM are a distraction at
best, a waste of the time and effort of otherwise reasonable individuals at
worst. What you need to do about QM, they say, is just learn it from some
undergraduate-level textbook ---which one, it doesn't matter, since the topic
has been carved in stone long ago and now it is just an exceedingly
straightforward application of linear algebra. Then you apply it to whatever
practical application gets you the most funding, money, h-index or just the
fastest path to tenure if you haven't got it yet, and you are done. If you do
not care about money, stability or prestige, you can choose the topic to which
you want to apply the finished theory known as QM on the basis of the common
interest of your department, of the scientific community, of humankind, or of
the whole biosphere, whichever group you consider to be the worthy beneficiary
of your efforts.

What is also remarkable is that the advocates of this position not only
decline to work in such an irrelevant enterprise as the foundations of QM
themselves (something that falls entirely within the bounds of their
scientific and personal freedom), but they very frequently add (and they do
that with vigor) that you shouldn't waste your time in that either\ldots or
something to that effect. Maybe this intensity of purpose is what has caused
the whole viewpoint to be sometimes dubbed ``\emph{Shut up and calculate!}'';
a motto which has been attributed to Feynman but that seems to have been
actually coined by \cite{Mermin2004} (even if he is not sure about
it).\footnote{\label{foot:Feynman_compute} In fact, although Feynman maybe
didn't say it, he was indeed close. When he is discussing the measurement
rules for the double-slit experiment in his famous Lectures, he writes
\citep[p.~10-1]{Feynman1963}: ``So at the present time we must limit ourselves
to computing probabilities.''}

The proponents of the shut-up-and-calculate approach to the understanding of
QM often perform this pastoral work (of trying to save as many lost souls as
possible from the bleak fate of irrelevance) wearing the dignified robes of
pragmatism, commitment to progress, clarity, and orthodoxy. Also, among their
ranks, we can find celebrated scientists and deep thinkers that have achieved
significant advances in so many fields, which makes their claims even more
appealing.

I don't know what the experience of the reader has been, but I have met this
friction, this headwind, this resistance, frequently in my (short) career.
When you start studying, you are young (typically), everything is new (by
definition), and every great scientist you read and meet is more a semi-god
than a human to you. Hence, if you happen to encounter the
shut-up-and-calculate command more often than the opposite in your early days
as a proto-scientist, you might buy it and stop asking questions about the
foundations of QM\ldots to yourself and to others as well. I know because I
bought it myself.

Fortunately, I have also been gifted (or cursed!) with an innate and
relentless curiosity and a lack of respect for authority figures. Hence, along
the years, I have read more, I have grown a healthy distrust, and I have come
to the conclusion that the shut-up-and-calculate advice is a \emph{very bad
one}. I think that it is just obvious that QM is a conceptual mess at the
moment, with many respectable and clever scientists talking past each other,
and I believe that acting as if the problem doesn't exist can only lead to
stagnation and to a delay in scientific progress. It is OK to close a topic if
it has been completely (or mostly) understood, but it is dangerous and
potentially tragic to close it too soon. I think that recommending young
scientists (or not-so-young ones) to shut up and calculate is not pragmatic,
does not help progress, does not foster clarity (it rather fosters quite the
opposite), and it is indeed not the orthodoxy.

As I said, many voices agree with me on this, so I will not tell anything new
here. My objective is not novelty but to join my voice to the choir that
advocates a deep investigation into the foundations and conceptual structure
of QM. In this way, I harbor the hope that the chances of you meeting a voice
of this second group (specially at the beginnings of your career) are slightly
increased by the existence of this document. Additionally, I will use the
opportunity to collect some interesting views on the controversy, and to put
my own thoughts in order.

Since I need a great deal of silence to concentrate (specially when I am
dealing with complicated topics) I have chosen to entitle these notes
``\emph{Shut up and let me think!}'' I think that this exclamation captures
the situation as well as my feelings about it quite
accurately.\footnote{\label{foot:contemplate} I was delighted to find that
very recently a similarly positive motto was coined by \cite{Hardy2010} to
counter the negative anti-rationalistic ``Shut up and calculate!''. They are
more polite than me, and they chose ``Shut up and contemplate!''.}

\section{Shut up and let me think!}
\label{sec:shut_up}

The position that I introduced in the previous section that defends the lack
of a necessity to investigate on the foundations of QM takes (of course)
different forms, it comes from different angles and it is stated with
different degrees of explicitness and intensity, depending on the personality
and beliefs of the person that expresses it. In this section, I will collect
some quotations\footnote{\label{foot:quotations} When you quote somebody, it
is easy to read something which is not actually said there, to misrepresent,
to forget the context, etc. I have tried not to make these mistakes, but given
the large number of quotations that I will mention, the odds are that I will
be putting the wrong words in somebody's mouth. For those researchers that are
still around, I would love to know if I have misquoted you, and I will
promptly include the appropriate addenda to this document if you show me my
error. For those that are no longer among us, such as Prof. Feynman, I
apologize in advance and I will also accept suggestions from anyone that knows
their actual thoughts better than I do.} and I will try to group them
according to a number of basic statements that are recurrent in the
literature, in conference talks, in corridor conversations and in the minds of
our fellow scientists.

Since I think that all these arguments are (at least partially) wrong, I will
also provide my personal answers to each of them.

I will begin some of the sections with a quote not because I want you to see
that a very important guy (or gal) agrees with me ---remember my lack of
respect in authority---, but because I think that they have (approximately)
said what I think (or the opposite) with a style that is typically much better
than mine.

\subsection{QM is completely understood}
\label{subsec:understood}

\begin{flushright}
\parbox[h]{5in}{{\small
What is the first business of one who practices philosophy? To get rid 
of self-conceit. For it is impossible for anyone to begin to learn that which 
he thinks he already knows.
\begin{flushright}
\vspace{-23.5pt}
--- Epictetus, Discourses, 101 AD
\end{flushright}
}}
\end{flushright}

\begin{flushright}
\parbox[h]{5in}{{\small
In this workshop we are venturing into a smoky area of science where
nobody knows what the real truth is. Such fields are always dominated by the
compensation phenomenon: supreme self-confidence takes the place of rational
arguments.
\begin{flushright}
\vspace{-23.5pt}
--- \cite{Jaynes1990a}
\end{flushright}
}}
\end{flushright}

This argument denies the premise by basically stating that QM is now
completely understood, and it has probably been so since a long time ago.
Nothing therefore needs to be done, and all the talk about the
``interpretations'', the ``measurement problem'', etc. is not physics but
philosophy. The proponents of this view enrich their argument with reasons and
sub-arguments supporting it, but I will deal with them later. In this section
I will mainly concentrate on the core assertion that everything about the
meaning and interpretation of QM has been solved already, and I will respond
to that.

The first great scientist that I want to mention that suggests everything is
understood in QM is Richard P. Feynman. In the very first page of the 3rd
volume of his awesome Lectures \citep{Feynman1963}, which is dedicated to 
---what else--- QM, he writes:

\begin{quote}
{\small The gradual accumulation of information about atomic and small-scale
behavior during the first quarter of this century, which gave some indications 
about how small things do behave, produced an increasing confusion which was
finally resolved in 1926 and 1927 by Schr\"odinger, Heisenberg, and Born. They
finally obtained a consistent description of the behavior of matter on a small
scale.}
\end{quote}

The same paragraph appears in p.~116 of his ``Six easy pieces'', and similarly
in \citep[p.~128]{Feynman1965} he writes:

\begin{quote}
{\small Electrons, when they were first discovered, behaved exactly like
particles or bullets, very simply. Further research showed, from electron
diffraction experiments for example, that they behaved like waves. As time
went on there was a growing confusion about how these things really behaved
---waves or particles, particles or waves? Everything looked like both. This
growing confusion was resolved in 1925 or 1926 with the advent of the correct
equations for quantum mechanics.}
\end{quote}

The funny thing is that, although Feynman seemed to think that the confusion
was ``resolved'' with a ``consistent description'' and the ``correct
equations'' for QM such a long time ago as in 1926-1927, he is the same person
that famously wrote (in the next page to the second quote above!):

\begin{quote}
{\small I think I can safely say that nobody understands quantum mechanics.}
\end{quote}

Also, in \citep{Feynman1982}, we can read:

\begin{quote}
{\small We always have had a great deal of difficulty in understanding the
world view that quantum mechanics represents. At least I do, because I'm an
old enough man that I haven't got to the point that this stuff is obvious to
me. Okay, I still get nervous with it. And therefore, some of the younger
students\ldots you know how it always is, every new idea, it takes a
generation or two until it becomes obvious that there's no real problem. It
has not yet become obvious to me that there's no real problem. I cannot define
the real problem, therefore I suspect there's no real problem, but I'm not
sure there's no real problem.}
\end{quote}

So it seems that in Feynman's mind the fact that all controversies have been
solved coexisted with a certain, undefined feeling that something was maybe
amiss. In my opinion, as every other human being, Feynman was perfectly
capable to hold contradictory thoughts in his mind at the same time, and
nevertheless display a functional and fruitful behavior. So probably at some
level he thought that QM presented no problems (hence the tone of the first
two quotes), but in other level he knew that something was not quite right
about it. Since he was mostly a pragmatist, he listened more to the first
voice than to the second and just went on to produce some of the grandest
achievements in modern physics, leaving that unresolved uneasiness in the back
of his mind to be dealt with later ---or never. Being an extraordinarily
honest thinker as he was, he did not pretend that the uneasiness did not
exist. He just preferred to advance without looking too much at it, or maybe
he found more interesting problems to look at than this one. However, it seems
clear that the possibility that everything relevant about the foundations of
QM was not in fact understood yet did have a place in his thoughts, as it also
floated in many conversations I have had with theoretical physicists, and as
the reader has also probably found in her travels.

A much stronger advocate of the no-problem-whatsoever view is another great
and respected scientist who is responsible of many important advances in
statistical mechanics: Nico van Kampen. In his letter entitled ``The scandal
of quantum mechanics'' \citep{VanKampen2008}, he attacks a paper by
\cite{Nikolic2008} about Bohmian mechanics ---one of the strongest proposals
in my opinion to solve the conceptual problems of ``orthodox''
shut-up-and-calculate QM---, and he writes paragraphs such as:

\begin{quote}
{\small The article by Nikoli\'c with its catchy title is a reminder of the
scandalous fact that eighty years after the development of quantum mechanics
the literature is still swamped by voluminous discussions about what is called
its ``interpretation''. Actually quantum mechanics provides a complete and
adequate description of the observed physical phenomena on the atomic scale.
What else can one wish?}
\end{quote}

Slightly after that, he proceeds to provide a two-paragraphs summary of what
he sees as the complete solution to the measurement problem:

\begin{quote}
{\small The solution of the measurement problem is twofold. First, any 
observation or measurement requires a macroscopic measuring apparatus. A 
macroscopic object is also governed by quantum mechanics, but has a large 
number of constituents, so that each macroscopic state is a combination of an 
enormous number of quantum mechanical eigenstates. As a consequence the 
quantum mechanical interference terms between two macroscopic states virtually 
cancel and only probabilities survive. That is the explanation why our 
familiar macroscopic physics, concerned with billiard balls, deals with 
probabilities rather than probability amplitudes.

[\ldots]

Second, in order that a macroscopic apparatus can be influenced by the 
presence of a microscopic event it has to be prepared in a metastable initial 
state ---think of the Wilson camera and the Geiger counter. The microscopic 
event triggers a macroscopically visible transition into the stable state. Of 
course this is irreversible and is accompanied by a thermodynamic increase of 
entropy.}
\end{quote}

And he concludes:

\begin{quote}
{\small This is the physics as determined by quantum mechanics. The scandal is 
that there are still many articles, discussions, and textbooks, which 
advertise various interpretations and philosophical profundities.}
\end{quote}

I tend to agree with van Kampen that the measurement problem is the
cornerstone of the conceptual issues that plague the ``orthodox'' accounts of
QM, and I also like his proposal very much. After giving a lot of thought to
it, and after comparing it to the most promising alternatives, I might even
decide that it is the ``right answer'' as far as I am concerned. However I
fail to see how the existence of his proposal is enough to dismiss the vast
literature that explores these topics ---and that works on the basis that
\emph{there is in fact a problem}--- as ``philosophical profundities''.

A specially poignant example of the literature I refer to is the excellent
work by \cite{Allahverdyan2013}, which, to my understanding, puts van Kampen's
two-paragraph summary of the solution to the measurement problem on solid and
concrete mathematical grounds (and which I also found very interesting). There
is a funny combination of facts about this very thorough and insightful
report:

\begin{itemize}

\item It was published online on November 14, 2012 ---that is, some years 
after the publication of the mentioned critique by \cite{VanKampen2008}.

\item \cite{VanKampen2008} cites no similar in-depth study to support his 
schematic account. He just treats it as an evident truth.

\item The article by \cite{Allahverdyan2013} is 166-pages long ---somewhat 
long for an evident truth.

\item The authors dedicate their work to ``\emph{our teachers and inspirers 
Nico G. van Kampen and Albert Messiah}''.

\item Far from dismissing all the literature concerning so-called 
``interpretations'' of QM as ``advertisements of philosophical profundities'', 
they write sentences such as the following:

\begin{quote}
{\small In spite of a century of progress and success, quantum mechanics still 
gives rise to passionate discussions about its interpretation. Understanding 
quantum measurements is an important issue in this respect, since measurements 
are a privileged means to grasp the microscopic physical quantities.

[\ldots]

The problem was thus formulated as a mathematical contradiction: the 
Schr\"o\-din\-ger equation and the projection postulate of von Neumann are 
incompatible. Since then, many theorists have worked out models of quantum 
measurements, with the aim of understanding not merely the dynamics of such 
processes, but in particular solving the so-called measurement problem. This 
problem is raised by a conceptual contrast between quantum theory, which is 
\emph{irreducibly probabilistic}, and our macroscopic experience, in which an 
\emph{individual process results in a well defined outcome}.

[\ldots]

The challenge has remained to fully explain how this property emerges, 
ideally without introducing new ingredients, that is, from the mere laws of 
quantum mechanics alone. Many authors have tackled this deep problem of 
measurements with the help of models so as to get insight on the 
interpretation of quantum mechanics.

[\ldots]

Few textbooks of quantum mechanics dwell upon questions of
interpretation or upon quantum measurements, in spite of their importance in
the comprehension of the theory. Generations of students have therefore
stumbled over the problem of measurement, before leaving it aside when they
pursued research work. Most physicists have never returned to it, considering
that it is not worth spending time on a problem which ``probably cannot be
solved'' and which has in practice little implication on physical predictions.
Such a fatalistic attitude has emerged after the efforts of the brightest
physicists, including Einstein, Bohr, de Broglie, von Neumann and Wigner,
failed to lead to a universally accepted solution or even viewpoint. [\ldots]
However, the measurement problem has never been forgotten, owing to its
intimate connection with the foundations of quantum mechanics, which it may
help to formulate more sharply, and owing to its philosophical implications.}
\end{quote}

Also, you can find a great paragraph full of references in their page~6 where
almost all ``philosophical profundities'' in the literature are considered to
be worth citing\ldots along with the names of those who dared to work in
this apparently non-existent problem.

\end{itemize}

One wonders how the words by \cite{VanKampen2008} can be compatible with the
enormous effort done by his disciples to clarify the whole issue. If
everything was already solved by 2008, then what is the point of writing a
very detailed, 166-page solution in 2012?

The question is of course rhetorical. The answer is that \emph{nothing was
solved by 2008, and it is probably not solved by now either}. The many
interesting works in the literature about ``interpretations'' of QM and the
measurement problem are not ``advertisements of philosophical profundities''
but serious attempts to make sense of a very deep conceptual problem, and the
many great scientists that have devoted time and effort to think about this
are not delusional and they have not forgotten their pragmatism at home. One
might wish that there is no problem with QM, but anyone that has her eyes open
sees that in fact \emph{there is} ---the belief that everything is OK is just
wishful thinking.

Of course, maybe Bohmian mechanics is not the solution and maybe van Kampen is
right about that. But even if that were the case it doesn't mean that there is
nothing to be solved to begin with. Maybe van Kampen's proposal is not the
solution for that matter either.

The excellent work by \cite{Allahverdyan2013} is not the only one we can
mention that explores this very interesting issue. In fact, tens of them are
published in the \href{http://arxiv.org/archive/quant-ph}{Quantum Physics
section} of the \href{http://arxiv.org/}{arXiv} every month, according to my
experience [see, for very recent examples, the pre-print \citep{Hobson2013},
and an answer to it \citep{Kastner2013} in the same week]. It would be
preposterous to make an exhaustive list here, so let me quasi-randomly cite
some very recent works by well-known scientists, most of them in very renowned
journals:
\citep{Bassi2013,Colbeck2012,Pusey2012,Colbeck2013,Froehlich2012,Kochen2013,Lewis2012,Pfister2013,Price2013,Saunders2010,Streltsov2013}.
These articles contain many questions that are still partially unanswered as
well as the heroic attempts of their authors to solve them. I will not discuss
the open issues here (I merely want to make clear that \emph{there are in fact
many open issues}), but I recommend any reader interested in ---and puzzled
by!--- the foundations of QM to enjoy a deep dive in any of these works to
begin mapping the landscape.

Another interesting reading that could help gain some perspective about the
lack of consensus in the field is the poll conducted by
\cite{Schlosshauer2013} to 33 participants (27 of which stated their main
academic affiliation as physics, 5 as philosophy, and 3 as mathematics) in the
conference ``Quantum physics and the nature of reality'', held in July 2011 at
the International Academy Traunkirchen, Austria. The authors asked the
conference participants 16 multiple-choice questions covering the main issues
and open problems in the foundations of QM. Multiple answers per question were
allowed to be checked because they were often non-exclusive. The results of
the poll are illuminating, and the authors conclude that:

\begin{quote}
{\small Quantum theory is based on a clear mathematical apparatus, has
enormous significance for the natural sciences, enjoys phenomenal predictive
success, and plays a critical role in modern technological developments. Yet,
nearly 90 years after the theory's development, there is still no consensus in
the scientific community regarding the interpretation of the theory's
foundational building blocks. Our poll is an urgent reminder of this peculiar
situation.}
\end{quote}

\begin{figure}[!t]
\begin{center}
\includegraphics[scale=0.35]{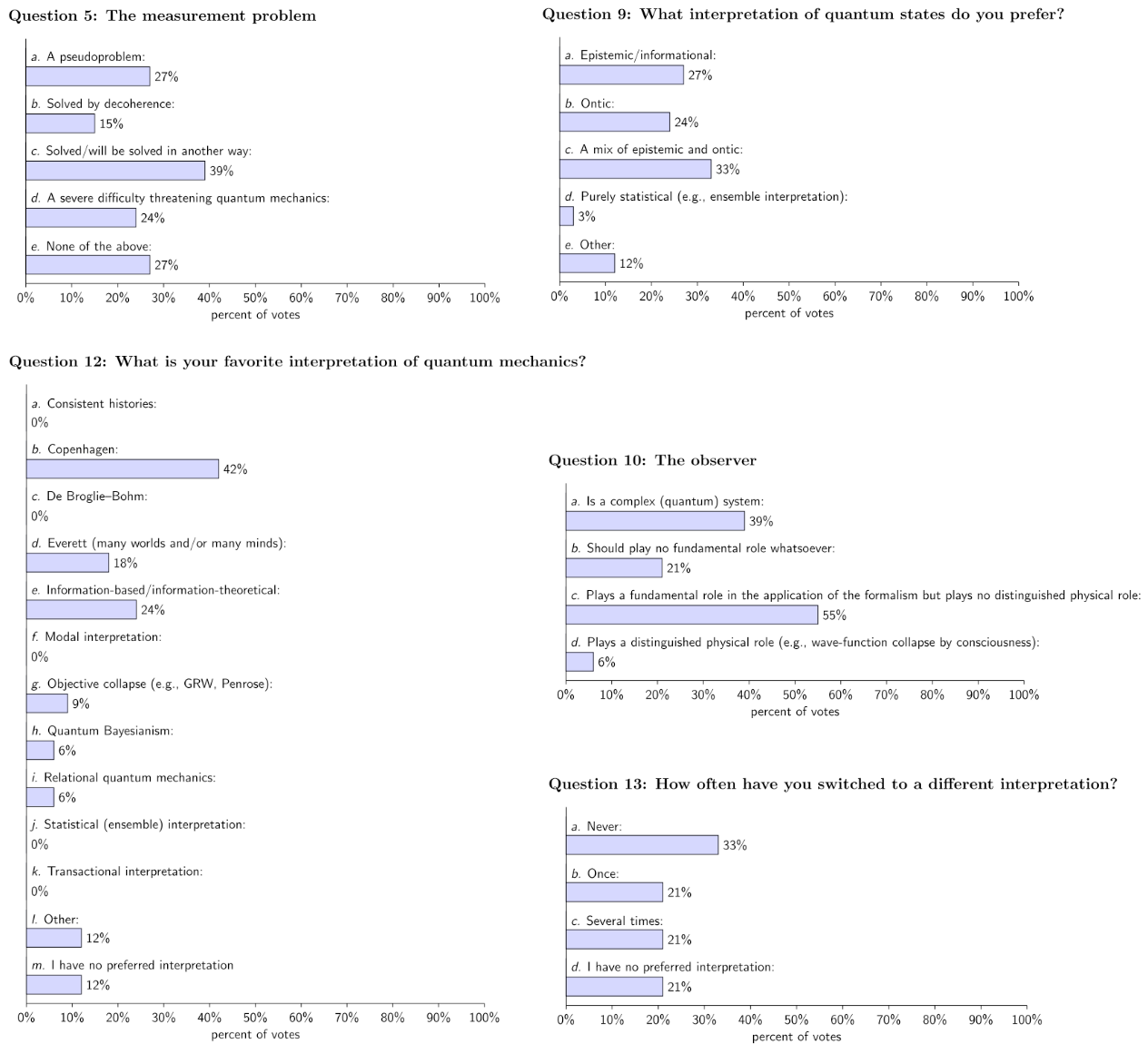}
\caption{\label{fig:poll}{\small Some results of the poll about foundational
attitudes toward QM conducted by \cite{Schlosshauer2013} as described in the
text. Used with kind permission by the authors.}}
\end{center}
\end{figure}

All the asked questions are very interesting, but I have taken the liberty to
select some of my favorite ones in fig.~\ref{fig:poll}. (I especially love the 
one related to changing your mind about which is the correct interpretation!)

Of course, scientific issues are not decided democratically, and moreover I
have already expressed my lack of respect for authority ---which I won't
withdraw. Nevertheless, it seems to me that an Ockham's razor type argument
leads to conclude that it is more likely that unsolved issues remain and those
few who deny it are victims of the ``compensation phenomenon'' mentioned in
Jaynes' quote at the beginning of this section, than it is that tens
---hundreds? thousands?--- of apparently clever and rational scientists of
different countries, of different ages, some of which don't even know each
other, are all investigating a no-problem.

Since this kind of plausibility arguments are never definitive, I must admit
that it is indeed possible that van Kampen's view is correct and all the
researchers that are pursuing alternative lines just don't know it yet. Still,
it seems that this view has been carefully formalized only in 2012
\citep{Allahverdyan2013} ---so it would appear that we need to grant a little
more time to the unconverted if we want to be fair. Even more! The same
argument can be used to admit the possibility that the many solid scientists
that claim that, say, Bohmian mechanics \citep{Durr2009} is the ultimate
solution to all of QM's conceptual problems be right and all those that don't
see it this way are just temporarily wrong. Or the many worlds interpretation
\citep{Saunders2010,Wallace2012}, or the consistent histories approach
\citep{Griffiths2003}, or quantum Bayesianism \citep{Fuchs2000}, or collapse
(also ``dynamical reduction'') models \citep{Bassi2003}, or \ldots

I think that the situation can be appropriately grasped if we admit to define
the sentences ``there is a problem in a theory'' or ``theory X is understood''
as referring not so much to physical reality or to the theory itself, but to
the scientific community that tinkers with it. Thus, a theory will have a
problem when a significant mass of intelligent scientists feels it has one,
and a theory will be completely understood when the vast majority of the
scientists feels it is so. I think I have proved that QM \emph{does have
problems and it is not completely understood} according to this point of view.
In sec.~\ref{subsubsec:classical}, I will show that there exist theories, such
as special relativity, which have no problems and are completely understood in
this sense ---just QM is not one of them.

Before closing this section, let me quote the words in
\citep[p.~87]{Streater2007}, who traces all the alleged wasted effort to a
failure in understanding that the ``EPR paradox'' is no ``paradox'' at all,
and mostly agrees with van Kampen on the no-problem-whatsoever diagnostic
(even if I am not completely sure that both propose the same solution):

\begin{quote}
{\small The quantum paradox community is, even after all these years and all
the rebuttals, still on the increase. In has spawned larger enterprises, such
as Bohmian mechanics, Nelson dynamics, geometro-stochastic dynamics,
quantum-state reduction, and the quantum theories of the brain of Penrose and
of Stapp. These are lost causes too, and fall down with the rebuttal of the
spooky action at a distance claimed to inspire them.}
\end{quote}

Again, we have to choose between believing that Streater knows the solution
but so many researchers don't, or believing that no complete solution has been
provided yet and it is Streater who doesn't realize that this is so.

To me, the clear lack of consensus expressed in the following final quotes and
epitomized in the poll I mentioned before is a clear sign that the second
option is much more likely than the first:

\begin{quote}
{\small We do not claim to offer any genuinely new or original thoughts on 
quantum mechanics. However, we have made the experience that there is still a 
fair amount of confusion about the deeper meaning of this theory --even among 
professional physicists. \citep{Froehlich2012}}
\end{quote}

\begin{quote}
{\small While its applications have made quantum theory arguably the most 
successful theory in physics, its interpretation continues to be the subject 
of lively debate within the community of physicists and philosophers concerned 
with conceptual foundations. \citep{Healey2012}}
\end{quote}

\begin{quote}
{\small Almost a century after the mathematical formulation of quantum 
mechanics, there is still no consensus on the interpretation of the theory.
\citep{Kochen2013}}
\end{quote}

Even some crazy scientists, such as \cite[chap.~11]{Carroll2010}, are brave
enough to call the whole business of interpreting QM ``respectable''! Do not
take my word for it. Read, read:

\begin{quote}
{\small There is a respectable field of intellectual endeavor, occupying the
time of a substantial number of talented scientists and philosophers, that
goes under the name of ``interpretations of quantum mechanics''. A century
ago, there was no such field as ``interpretations of classical mechanics''
---classical mechanics is perfectly straightforward to interpret. \emph{We're
still not sure what is the best way to think and talk about quantum
mechanics.} [my emphasis]}
\end{quote}

Despite these encouraging words, the paper by \cite{Englert2013}, which was
published in the arXiv curiously the same day as the first version of this
manuscript, is the living proof that the ``Shut up and calculate!'' school is
strong and thriving. We could delight in many of its paragraphs, but it
actually suffices to read the unusually short abstract:

\begin{quote}
{\small Quantum theory is a well-defined local theory with a clear 
interpretation. No ``measurement problem'' or any other foundational matters 
are waiting to be settled.}
\end{quote}

\subsubsection{QM is completely understood by omission}
\label{subsubsec:understood_omission}

\begin{flushright}
\parbox[h]{5in}{{\small
Since Bohmian mechanics is so simple and straightforward, only one 
criticism remains: there must be something wrong with Bohmian mechanics, 
otherwise it would be taught. And as a consequence, Bohmian mechanics is not
taught because there must be something wrong with it, otherwise it would be 
taught.
\begin{flushright}
\vspace{-23.5pt}
--- \cite{Durr2009}
\end{flushright}
}}
\end{flushright}

Despite the strong opinions of van Kampen and Streater discussed in the
previous section, and despite many similar public views of other scientists on
the topic, it is my impression that the sheer force of the facts behind the
lack of consensus that I think have I proved prevents most people from
thinking that QM is a finished theory ---or if they think it, they don't say
it. There is, however, a subtler way of suggesting the same thing: \emph{by
omission}.

If you write a text and you never say that QM is completely understood, but
you don't say the contrary either (or maybe you mention something about it
briefly and \emph{en passant}), the same viewpoint can be conveyed.

Of course, it would be inefficient to make a detailed comment about the
conceptual mess that pervades the foundations of QM in every paper that uses
the theory or talks about it. What one should expect, nevertheless, is that
such a discussion \emph{would be provided} in textbooks, lectures and other
occasions intended to introduce the topic to newcomers. Failing to mention
something in an account of such aim amounts to tacitly stating that the
omitted point is not important to get a solid knowledge of the subject.

Since no interpretation is settled as ``the right one'', the author of the
introductory material has to pick one and talk about it as if it were the only
game in town. The Copenhagen view, which is typically seen as the one that
most closely corresponds to the ``Shut up and calculate!''~motto, is chosen to
play this role in almost all cases. Therefore, generation after generation of
students begin their careers with the implicit impression that this is how
\emph{things are} ---as in the most conservative families, the kids are taught
to shut up when they are still young and obedient. Only this time they are not
told so explicitly (at least not always). Even the proponents of the
shut-up-and-calculate approach realize from time to time that telling people
not to think may sound a little unscientific.

I have found this omission not only in my own formation as a physicist but
also in online university courses \citep{Clark2013}, as well as in most of the
otherwise fantastic textbooks about QM that are typically used in such
courses. Let me mention only a few examples:
\citep{Cohen-Tannoudji1977,Cohen-Tannoudji1977a,Messiah1961,Messiah1961a,Landau1991,Newton2002,Sakurai1994,Shankar1994}.
Some other textbooks do assign importance to interpretative matters, but they
present them as more or less solved according to one or another proposal of
the authors' liking \citep{Ballentine1998}. In yet some other textbooks,
foundational issues are discussed and the different alternatives presented,
but they are often included into the more ``advanced topics'' and not as
concerns affecting the conceptual stability of the whole edifice
\citep{Auletta2009}.

Let me quote \cite[chap.~11]{Carroll2010} again so you can see that I am not
the only crazy fellow saying this:

\begin{quote}
{\small But the history is less crucial to our present purposes than the
status of the Copenhagen view as what is enshrined in textbooks as the
standard picture. Every physicist learns this first, and then gets to
contemplate other alternatives (or choose not to, as the case may be).}
\end{quote}

It is my content here that what is needed is to \emph{begin} any introductory
course (or textbook) in QM with a warning: ``\emph{Be careful! The theory that
I am going to describe to you next is extremely predictive when one applies
its pseudo-rules to concrete experiments creatively. However, our conceptual
understanding of it is still fuzzy and preliminary. How to make it solid I do
not know, and it is by the way one of the most fascinating open scientific
questions that you could decide to tackle once you graduate.}'' This is not
usually done in my experience, and boy would it have been useful to me some
years ago!

\subsection{OK, maybe we don't understand it, but that is not our job}
\label{subsec:understanding}

\begin{flushright}
\parbox[h]{5in}{{\small
What worries me about religion is that it teaches people to be satisfied 
with not understanding.
\begin{flushright}
\vspace{-9pt}
--- Richard Dawkins, \href{http://www.youtube.com/watch?v=2gTYFolrpNU}{God Under The Microscope}, BBC, 1996
\end{flushright}
}}
\end{flushright}

\begin{flushright}
\parbox[h]{5in}{{\small
I think I can safely say that nobody understands quantum mechanics.
\begin{flushright}
\vspace{-9pt}
--- \cite{Feynman1965}
\end{flushright}
}}
\end{flushright}

It is not yet clear to me what ``understanding'' a theory really is, but it
seems a sure bet that it should require that one knows how the different
technical terms that appear within it are related to concrete experiments;
that one knows how to apply the theory in every practical case without
hesitation; that one knows how to relate the theory to other theories properly
(think about statistical mechanics, classical mechanics, or relativity in the
case of QM); that one feels at ease and satisfied when using the theory and
not perpetually asking oneself whether or not the calculations make sense,
whether or not the next analytical step that one is planning to take is valid;
etc. For me, ``understanding'' is something that can be predicated about the
network of concepts that make up the theory, as well as its embedding in the
larger network that comprises the whole worldview of the individual that is
trying to ``understand''. If something is amiss, if the network seems 
unstable, this translates into a lack of security when reasoning about the 
theory and its consequences.

But it is not my purpose to perfectly define what ``understanding'' means. It
suffices to me to appeal to the popular use of the word in everyday life to
make my point. In this, I agree with \cite[p.~8-2]{Feynman1963a}:

\begin{quote}
{\small We cannot define \emph{anything} precisely. If we attempt to, we get
into that paralysis of thought that comes to philosophers, who sit opposite to
each other, one saying to the other: ``You don't know what you are talking
about!''. The second one says: ``What do you mean by \emph{know}? What do you
mean by \emph{talking}? What do you mean by \emph{you}?'', and so on.}
\end{quote}

In fact, Feynman didn't define the word and yet he used it in the quote that
opens this section. The important thing is that, whatever Feynman felt that
``understanding'' is, he claimed that ``nobody understands QM'' ---not even
himself! On the other hand, although this seems to have produced certain
restlessness in his mind according to the quotes in
sec.~\ref{subsec:understood}, he didn't think that this lack of understanding
was in any way incompatible with the fact (for him it was a fact) that all the
confusion regarding the behavior of matter at the microscopic level was
``resolved'' with a ``consistent description'' and the ``correct equations''
for QM such a long time ago as in 1926-1927 (as we have already quoted him
saying).

It is the second claim that I challenge. I don't think that you can say that
anything has been ``resolved'' or that your description is ``consistent''
unless you understand your theory. Some voices that agree with Feynman that QM
is not yet understood circumvent the problem by saying in one way or another
that understanding is unnecessary, that physics is not concerned with it, that
it is not our job. It is clear from my position that I just don't think this
is true. In my opinion, every scientist needs to understand everything that is
related to her field of expertise ---and physicists are not the exception. I
also think that the assertion that ``understanding is not the job of physics''
is an example of the normative mode in which people tend to enter when they
feel that something is deeply wrong but they don't want to face the problem
head on. In what follows, we will meet some additional unjustified rules that
seem invented ad hoc to protect QM from inquiry, from analysis, from repair.
But let me quote here what \cite{Svozil2012} has to say about a quote by
Feynman that we will meet again later:

\begin{quote}
{\small Feynman's advise to (young, as he seemed to have assumed that the
older ones are sufficiently brainwashed anyway) physicists that, while
``\ldots nobody understands quantum mechanics'', to stop thinking about these
issues ``\ldots if you can possibly avoid it. Because you will get `down the
drain', into a blind alley from which nobody has yet escaped. Nobody knows how
it can be like that.'' ---Even if one grants Feynman some rhetoric benefits,
the appeal to ``stop thinking'' is a truly remarkable advice from one of the
most popular scientists of his time!}
\end{quote}

Indeed! It reminds me so much of the layers of protection again rational
inquiry that religion has developed ---faith, irreducible mystery, being
``out-of-bounds'' of science, etc. [see \citep{Dennett2006} for an
illuminating account of these mechanisms]--- and yet it is supposed to be
physics! It seems to me that, once you admit that QM is not understood ---as
Feynman did---, you cannot elude your responsibility to understand it by
producing ad hoc rules about what can be investigated, what is out of bounds,
what you can think about and what not. In words of
\citep[chap.~11]{Carroll2010}:

\begin{quote}
{\small Most modern physicists deal with the problems of interpreting quantum 
mechanics through the age-old strategy of ``denial''. They know how the rules 
operate in cases of interest, they can put quantum mechanics to work in 
specific circumstances and achieve amazing agreement with experiment, and they 
don't want to be bothered with pesky questions about what it all means or 
whether the theory is perfectly well-defined.}
\end{quote}

But before I convince you ---yes, I am an optimist--- let us analyze some
arguments related to this overarching theme that I have dubbed ``understanding
is not our job''. First of all, two weaker versions of the idea.

\subsubsection{OK, maybe understanding is our job, but it is impossible}
\label{subsubsec:impossible}

\begin{flushright}
\parbox[h]{5in}{{\small
When a distinguished but elderly scientist states that something is 
possible, he is almost certainly right. When he states that something is 
impossible, he is very probably wrong.
\begin{flushright}
\vspace{-23.5pt}
--- The first of Arthur C. Clarke's three laws
\end{flushright}
}}
\end{flushright}

\begin{flushright}
\parbox[h]{5in}{{\small
Philosophers' Syndrome: mistaking a failure of the imagination for an 
insight into necessity.
\begin{flushright}
\vspace{-23.5pt}
--- \cite{Dennett1991}
\end{flushright}
}}
\end{flushright}

Feynman ---yes, he again--- wrote:

\begin{quote}
{\small So at the present time we must limit ourselves to computing
probabilities. We say ``at the present time'', but we suspect very strongly
that it is something that will be with us forever ---that it is impossible to
beat that puzzle--- that this is the way nature really \emph{is}.
\citep[p.~10-1]{Feynman1963}}
\end{quote}

And you probably met this objection or similar ones before: The mystery is
irreducible, the puzzle is eternal, it is impossible to crack the problem, and
so on and so forth. This is the central tenet of the Copenhagen 
interpretation, as we will see later.

I will not take too much time to discuss this anti-rationalistic argument. It
is too weak an adversary. Let me just mention that nobody has ever proved such
a thing about anything. So far, we humans have been able to slowly but
steadily understand more and more of nature's secrets, and it doesn't look to
me that QM will be special. In any case, if it were, the burden of the proof
is on those that say that understanding is impossible, not on us who have
plenty of examples of the contrary.

To be fair, I have to say that I don't believe that Feynman was
anti-rationalistic ---far from that! What I think is that, in his classic
energetic style, and somewhat uncomfortable because he himself knew that
something was wrong with the current understanding of QM, he mixed things a
little bit again in the above quote. I will discuss something more about this
soon in sec.~\ref{subsubsec:how_not_why}, but let me here briefly say that
believing that QM is ``the way nature really is'' is perfectly compatible with
the position that it is poorly understood and that something should be done
about that. Understanding QM is not finding a more fundamental way that
``nature really is'' and obtaining QM as a special case. (This might be
possible or not, and I am OK with both situations.) What is understanding to
me is (more or less) described at the beginning of
sec.~\ref{subsec:understanding}, and it is independent from that. Whether or
not there is an underlying theory below QM is not a puzzle, it is just a
question ---and one well within the boundaries of physics by the way!

\subsubsection{OK, maybe it is possible, but many geniuses have tried with no luck}
\label{subsubsec:luck}

\begin{flushright}
\parbox[h]{5in}{{\small
Quit now, you'll never make it. If you disregard this advice, you'll be 
halfway there.
\begin{flushright}
\vspace{-23.5pt}
--- David Zucker, (North) American film director
\end{flushright}
}}
\end{flushright}

\begin{flushright}
\parbox[h]{5in}{{\small
If at first you don't succeed, try, try, again. Then quit. There's no use 
being a damn fool about it.
\begin{flushright}
\vspace{-23.5pt}
--- W. C. Fields, (North) American comedian
\end{flushright}
}}
\end{flushright}

Taking again the risk of misreading Feynman's words, let us look at two more
quotes by him (one of which we already met):

\begin{quote}
{\small One might still like to ask: ``How does it work? What is the machinery
behind the law?'' No one has found the machinery behind the law. No one can
``explain'' any more than we have just ``explained''. No one will give you any
deeper representation of the situation. We have no ideas about a more basic
mechanism from which these results can be deduced.
\citep[p.~1-10]{Feynman1963}}
\end{quote}

\begin{quote}
{\small Do not keep saying to yourself, if you can possible avoid it, ``But
how can it be like that?'' because you will get `down the drain', into a blind
alley from which nobody has escaped. Nobody knows how it can be like that.
\citep[p.~129]{Feynman1965}}
\end{quote}

Probably mixing again the ``whys'' with the ``hows'' (see
sec.~\ref{subsubsec:how_not_why}), and also mixing the lack of a proper
understanding of QM with the lack of an underlying theory from which QM can be
derived, these two quotes nevertheless remind us of a plausibility argument
that you have probably heard many times: ``A lot of clever guys have spent
decades trying to solve this problem. Therefore, it is either impossible or
terribly difficult. If you decide to go into it you will most probably lose
your mind and spend your last days as a hermit in a dirty hut.''

Well, I don't deny some degree of truth to this line of reasoning. The problem
of understanding QM seems indeed difficult ---but it is also very important.
What remains is a personal (not a scientific) decision: Do you choose to take
these facts as a warning, a ``Beware of the dog!'' sign, or you choose to take
them as a challenge, as an opportunity?

\subsubsection{We need to find out the how, not the why}
\label{subsubsec:how_not_why}

\begin{flushright}
\parbox[h]{5in}{{\small
Teleology is like a mistress to a biologist: he cannot live without her 
but he's unwilling to be seen with her in public.
\begin{flushright}
\vspace{-23.5pt}
--- J. B. S. Haldane
\end{flushright}
}}
\end{flushright}

The first more complex idea about understanding QM that floats around (and
which is entangled with Feynman's statements in the two previous sections)
will not take me too much to discuss because I basically agree with it. It
says that physics does not need to worry about ``whys'', but only about
``hows''. This is clearly captured in the following quotes by, who else,
Feynman:

\begin{quote}
{\small Newton was originally asked about his theory ---`But it doesn't mean
anything ---it doesn't tell us anything.' He said, `It tells you \emph{how} it
moves. That should be enough. I have told you how it moves, not why.' But
people often are unsatisfied without a mechanism\ldots
\citep[p.~37]{Feynman1965}}
\end{quote}

\begin{quote}
{\small So there is no model of the theory of gravitation today, other than
the mathematical from. [\ldots] Every one of our laws is a purely mathematical
statement in rather complex and abstruse mathematics. [\ldots] Why? I have not
the slightest idea. \citep[p.~39]{Feynman1965}}
\end{quote}

Although it might seem at first sight that this is another example of a norm
invented for the sake of eluding our job of understanding QM, this is not so
for an important reason: Basically, the systems which are undisputedly the
traditional domains of physics and whose characteristics are embodied in the
mathematical objects and equations that are physics' language \emph{do not 
have whys}. Atoms do not have whys, nor do photons, or polymers, or solids,
or stars ---that is, unless you believe in a God that has embedded his own
purpose into every system in the Universe; if you do, then the \emph{why} of
every physical system is God's why, and it is outside physics too.

It is not easy to define what exactly a ``why'' is [you can read the great
book by \cite{Deacon2012} if you feel like knowing more about it], but I can
briefly mention that it requires an end-directness, a final cause. This final
cause can be represented in the minds of agents as it is often the case with
humans, or it can be free-floating [in the words by \cite{Dennett1995}], like
the whys that help us make sense of the evolution of biological organisms. As
jokingly expressed by Haldane's quote at the beginning of this section, it
exists a heated debate in biology about whether or not the use of whys is
scientifically justified. It is disputed whether biological organisms actually
have whys, but it is unanimously accepted that they indeed \emph{appear} to
have them. This begs the question: ``Are whys within the domain of biology?''
In physics, however, the behavior of the studied systems is typically captured
by differential equations which are local in time and do not appear to have
any ``final cause'' built into them. Physical systems do not even
\emph{appear} to have whys. Hence the answer to the question ``Are whys within
the domain of physics?'' is clearly and simply \emph{no}.

The first quote by Feynman above suggests a different conception of what a
``why'' is. Feynman seems to think that a why is just a mechanism behind a
law. That is, another more fundamental law. In this sense, equilibrium
statistical mechanics would be thermodynamics' why. But also in this sense the
answer to the question ``Are whys within the domain of physics?'' is again
trivial: It is \emph{yes}. In fact, I claim that this kind of whys is such
that it would be more precise if we called them ``hows''. A mechanism, a law
behind the law, seems more a how to me than a why ---and that is why it falls
within the competences of physics, by the way.

In any case, neither one type of why nor the other are part of what I have
called ``understanding'' (see the beginning of sec.~\ref{subsec:understanding}
for a sketchy definition of the concept). Understanding is not providing a
final cause or a mechanism ---although both could be part of understanding in
particular cases that do present final causes or underlying mechanisms as
valid ingredients. Hence, it is true that whys (in the first sense) are
outside the scope of physics, but that doesn't mean that understanding is. As
I said, I believe that understanding is a task that no discipline can avoid.
If we use the second sense (a more fundamental description), then whys are
trivially part of a physicist's job. In such a case, our obligation is to
understand \emph{both} QM \emph{and} its hypothetical underlying mechanism.

\subsubsection{Philosophy and interpretation as dirty words}
\label{subsubsec:philosophy}

\begin{flushright}
\parbox[h]{5in}{{\small
There is no such thing as philosophy-free science; there is only science 
whose philosophical baggage is taken on board without examination.
\begin{flushright}
\vspace{-9pt}
--- \cite{Dennett1995}
\end{flushright}
}}
\end{flushright}

\begin{flushright}
\parbox[h]{5in}{{\small
All science is either physics or stamp collecting.
\begin{flushright}
\vspace{-23.5pt}
--- Ernest Rutherford
\end{flushright}
}}
\end{flushright}

One literary ornament that people sometimes use when trying to defend that
understanding QM (or any other theory for that matter) is not the job of the
physicists is to assign negative value to neutral terms such as ``philosophy''
or ``interpretation'' (``semantics'' is also used sometimes) and apply them to
any endeavor that they consider futile. We already met an example of this nice
practice in the quote by van Kampen in sec.~\ref{subsec:understood} where he
derogatorily calls most of the literature on the measurement problem and
Bohmian mechanics ``philosophical profundities''.

As we can read in the blog ``Backreaction'' \citep{Hossenfelder2013}, this is
not uncommon among physicists:

\begin{quote}
{\small I know a lot of physicists who use the word philosophy as an insult
\ldots}
\end{quote}

Physicists are typically clever guys ---I know because I am one!--- and, as
clever guys often do, they sometimes think that they are more clever than they
actually are. They do not say it directly, but they typically assign this
excess of intelligence to the discipline itself. ``The thing is not that I am
extraordinarily clever and handsome'', they say, ``but that physics is a
superior intellectual scheme to all the rest of ways of studying the world.
Since I happen to be a physicist, well, I guess that I am superior too, but I
don't like to say it aloud not to sound immodest.'' The quote by Rutherford at
the beginning of this section is an example of this humble attitude, and our
old friend Richard P. Feynman is also known for having expressed derogatory
opinions about philosophy and philosophers now and then. Here you have some
quotes from his very enjoyable biography ``Surely you're joking, Mr. Feynman''
\citep{Feynman1985}, but I am sure that you can find many more if you spend
10 minutes in Google:

\begin{quote}
{\small Another guy got up, and another, and I tell you I have never heard 
such ingenious different ways of looking at a brick before. And, just like it 
should in all stories about philosophers, it ended up in complete chaos. In 
all their previous discussions they hadn't even asked themselves whether such 
a simple object as a brick, much less an electron, is an ``essential 
object''.}
\end{quote}

\begin{quote}
{\small In the early fifties I suffered temporarily from a disease of middle 
age: I used to give philosophical talks about science\ldots}
\end{quote}

\begin{quote}
{\small Cornell had all kinds of departments that I didn't have much interest 
in. (That doesn't mean there was anything wrong with them; it's just that I 
didn't happen to have much interest in them.) There was domestic science, 
philosophy (the guys from this department were particularly inane), and there 
were the cultural things ---music and so on. There were quite a few people I 
did enjoy talking to, of course. In the math department there was Professor 
Kac and Professor Feller; in chemistry, Professor Calvin; and a great guy in 
the zoology department, Dr. Griffin, who found out that bats navigate by 
making echoes. But it was hard to find enough of these guys to talk to, and 
there was all this other stuff which I thought was low-level baloney. And 
Ithaca was a small town.}
\end{quote}

Notice in this last quote how he gracefully moves from ``that doesn't mean
there was anything wrong with them'' to ``low-level baloney'' in just one
paragraph.

We can also read the following in \citep[chap.~2]{Feynman1963a}:

\begin{quote}
{\small Philosophers, incidentally, say a great deal about what is absolutely 
necessary for science, and it is always, so far as one can see, rather naive, 
and probably wrong.}
\end{quote}

We can read him equating ``philosophical principles'' to feelings and personal 
likings in \citep[p.~57]{Feynman1965}:

\begin{quote}
{\small The next question is whether, when trying to guess a new law, we
should use the seat-of-the-pants feeling and philosophical principles ---`I
don't like the minimum principle', or `I do like the minimum principle', `I
don't like action at a distance', or `I do like action at a distance'.}
\end{quote}

Or we can check what he thought about philosophers' opinions regarding science
in \citep[p.~147]{Feynman1965}:

\begin{quote}
{\small A philosopher once said `It is necessary for the very existence of
science that the same conditions always produce the same results.' Well, they
do not [\ldots] What is necessary `for the very existence of science', and
what the characteristics of nature are, are not to be determined by pompous
preconditions, they are determined always by the material with which we work,
by nature herself. We look, and we see what we find, and we cannot say ahead
of time successfully what is going to look like. [\ldots] In fact it is
necessary for the very existence of science that minds exist which do not 
allow that nature must satisfy some preconceived conditions, like those of our
philosopher.}
\end{quote}

Incidentally, I agree with Feynman that we should avoid inventing norms about
``what science is'', ``what a theory is'', ``what we can talk about'', etc.
---I just don't think that philosophers are the only ones who concoct such ad
hoc rules. As I mentioned in sec.~\ref{subsubsec:how_not_why} and will expand
in sec.~\ref{subsec:rules}, physicists are also very fond of repeating 
statements in the spirit of ``It is necessary for the very existence of
science that the same conditions always produce the same results'', and which
seem to come from nowhere.

To be fair, I have to say that it is not only physicists who sometimes neglect
the usefulness of other disciplines. This is actually an epidemic.
Mathematicians tend to see physicists as careless and without rigor, while
physicists sometimes declare that mathematicians worry about unimportant
matters. Chemists usually look at physics' topics with skepticism (``OK, and
how many decades until we know if there is even a small probability that this
finds an application?''), while physicists are often heard saying that
chemists use the tools of physics without properly understanding them
---especially QM. A similar relationship is in place between chemists and
biochemists, between biochemists and biologists, between biologists and
psychologists, between psychologists and sociologists, and so on and so forth.

Coming back to the relationship between physics and philosophy, you probably
experienced the following situation which is also mentioned in the
``Backreaction'' blog \citep{Hossenfelder2013}:

\begin{quote}
{\small I've heard talks by philosophers about the ``issue'' of infinities in 
quantum field theory who had never heard of effective field theory. I've heard 
philosophers speaking about Einstein's ``hole argument'' who didn't know what 
a manifold is, and I've heard philosophers talking about laws of nature who 
didn't know what a Hamiltonian evolution is.}
\end{quote}

Well, yes\ldots I have been to very bad talks indeed ---and not all of them
were by philosophers. But on the other hand\ldots

\begin{quote}
{\small But on the other hand, I've met remarkably sharp philosophers with the 
ability to strip away excess baggage that physicists like to decorate their 
theories with, and go straight to the heart of the problem. 
\citep{Hossenfelder2013}}
\end{quote}

Exactly! So it seems that the solution is \emph{very simple}: stick to good
philosophy (physics, chemistry, sociology, \ldots), and forget the part of the
discipline that is low quality. Or, as Dennett more eloquently puts it, accept
\emph{Sturgeon's Law} and act consequently \citep[sec.~II.4]{Dennett2013}:

\begin{quote}
{\small \emph{Ninety percent of everything is crap.} Ninety percent of
experiments in molecular biology, 90 percent of poetry, 90 percent of
philosophy books, 90 percent of peer-reviewed articles in mathematics ---and
so forth--- is crap. Is that true? Well, maybe it's an exaggeration, but let's
agree that there is a lot of mediocre work done in every field. (Some
curmudgeons say it's more like 99 percent, but let's no get into that game.) A
good moral to draw from this observation is that when you want to criticize a
field, a genre, a discipline, an art form, \ldots \emph{don't waste your time
and ours hooting at the crap!} Go after the good stuff, or leave it alone.
This advice is often ignored by ideologues intent on destroying the reputation
of analytic philosophy, evolutionary psychology, sociology, cultural
anthropology, macroeconomics, plastic surgery, improvisational theater,
television sitcoms, philosophical theology, massage therapy, you name it.}
\end{quote}

For example, let us read how Hilary Putnam (a philosopher who many would argue
belongs to the 10\% of non-crap) helped a renowned physicist change his mind
about the need of a conceptual tidying-up in QM \citep{Putnam2005}:

\begin{quote}
{\small For myself, and for any other `scientific realist', the whole
so-called interpretation problem in connection with quantum mechanics is just
this: \emph{whether} we can understand quantum mechanics ---no, let me be
optimistic--- \emph{how} to understand quantum mechanics in a way that is
compatible with the anti-operationalist philosophy that I subscribed to in the
pages I just quoted, and that I have always subscribed to. But it took a long
time for physicists to admit that there is such a problem. I can tell you a
story about that. In 1962 I had a series of conversations with a world-famous
physicist (whom I will not identify by name). At the beginning, he insisted,
`You philosophers just \emph{think} there is a problem with understanding
quantum mechanics. We physicists have known better from Bohr on.' After I
forget how many discussions, we were sitting in a bar in Cambridge, and he
said to me, `You're right. You've convinced me there is a problem here; it's a
shame I can't take three months off and solve it.'

Fourteen years later, the same physicist and I were together at a conference
for a few days, and he opened his lecture at that conference (a lecture which
explained to a general audience the exciting new theories of quarks) by
saying, `There is no Copenhagen interpretation of quantum mechanics. Bohr
brainwashed a generation of physicists.' Evidently, he had undergone a
considerable change of outlook.}
\end{quote}

In the following quote, also by \cite{Putnam2010}, we can read some more
thoughts about the relationship between philosophy and science, we find a
clear exposition about the real nature of the so-called ``Copenhagen
interpretation'', and we learn that the physicist in the above story was the
Nobel Prize winner Murray Gell-Mann :

\begin{quote}
{\small That philosophy is not to be identified with science is not to deny
the intimate relation between science and philosophy. The positivist idea that
all science does is predict the observable results of experiments is still
popular with some scientists, but it always leads to the \emph{evasion} of
important foundational questions. For example, the recognition that there is a
problem of \emph{understanding} quantum mechanics, that is, a problem of
figuring out just how physical reality must be in order for our most
fundamental physical theory to work as successfully as it does, is becoming
more widespread, but that recognition was \emph{delayed} for decades by the
claim that something called the ``Copenhagen interpretation'' of Niels Bohr
had solved all the problems. Yet the ``Copenhagen interpretation'', in Bohr's
version, amounted only to the vague philosophical thesis that the human mind
couldn't possibly understand how the quantum universe was in itself and should
just confine itself to telling us how to use quantum mechanics to make
predictions \emph{stateable in the language of classical, that is to say,
non-quantum-mechanical, physics!} (in my lifetime, I first realized that the
``mood'' had changed when I heard Murray Gell-Mann say in a public lecture
sometime around 1975 ``There is no Copenhagen interpretation of quantum
mechanics. Bohr brainwashed a generation of physicists!'')

Only after physicists stopped being content to regard quantum mechanics as a
mere machine for making predictions and started taking seriously what this
theory actually \emph{means} could real progress be made. Today many new paths
for research have opened as a result: string theory, various theories of
quantum gravity, and ``spontaneous collapse'' theory are only the beginning of
quite a long list. And Bell's famous theorem, which has transformed our
understanding of the `measurement problem'. would never have been proved if
Bell had not had a deep but at the time highly unpopular interest in the
meaning of quantum mechanics.}
\end{quote}

To me the issue is clear. The position that physics can be done without appeal
to meaning, to interpretation, to ``philosophy'', is just untenable. The idea
that physics is only a little more than mathematics, that applying physical
theories to actual experiments is straightforward, that ``Shut up and
calculate!'' is a good advice, are nothing but expressions of wishful
thinking. You can close your eyes in the easiest cases and everything might
work well for a while; however, when the range of problems to which you want
to apply the theory is enlarged, when you want to relate the theory to other
neighboring theoretical schemes, when you embark in the construction of new
models of nature, or when you want to grasp the workings of the theory beyond
the blind mechanical implementation of its equations as if they were cooking
recipes,\ldots in all these situations, you might find that you need to take a
walk in the dreaded fields of ``philosophy''. At that moment, you may find the
insights by some philosophers helpful (some philosophers belonging to the
non-crappy 10\%, that goes without saying), and you will probably stop using
``philosophical'' as a dirty adjective.

Let me close this section with some quotes by scientists who have also 
realized that this is the case, so you can see that Murray Gell-Mann and I are
not the only ones to hold this opinion:

\begin{quote}
{\small Which view one adopts affects how one thinks about the theory at a 
fundamental level. \citep{Colbeck2012}}
\end{quote}

\begin{quote}
{\small How we look at a theory affects our judgment as to whether it is
mysterious or irrational on the one hand; or whether it is satisfactory and
reasonable on the other. Thus it affects the direction of our research
efforts; and \emph{a fortiori} their results. Indeed, whether we theorists can
ever again manage to get ahead of experiment will depend on how we choose to
look at things, because that determines the possible forms of the future
theories that will grow out of our present ones. One viewpoint may suggest
natural extensions of a theory, which cannot even be stated in terms of
another. What seems a paradox on one viewpoint may become a platitude on
another. \citep{Jaynes1990a}}
\end{quote}

\begin{quote}
{\small Because so many of the results do not seem to depend in a critical way 
on the choice of interpretation, some ``practical-minded'' physicists would 
like to dismiss the whole subject of interpretation as irrelevant. That 
attitude, however, is not justified, and a number of practical physicists have 
been led into unnecessary confusion and dispute because of inattention to the 
matters of interpretation that we have been discussing.
\citep[p.~239]{Ballentine1998}}
\end{quote}

\subsubsection{Understanding equals being intuitive, or being classical}
\label{subsubsec:classical}

\begin{flushright}
\parbox[h]{5in}{{\small
Brad DeLong, in the course of something completely different, suggests that 
the theory of relativity really isn't all that hard. At least, if your 
standard of comparison is quantum mechanics.
\begin{flushright}
\vspace{-23.5pt}
--- \cite{Carroll2011}
\end{flushright}
}}
\end{flushright}

Another way in which you can reject the need of understanding QM is by
claiming that, actually, what ``understanding'' \emph{is} is something
\emph{unnecessary} and henceforth of very little value.

\begin{figure}[!ht]
\begin{center}
\includegraphics[scale=0.3]{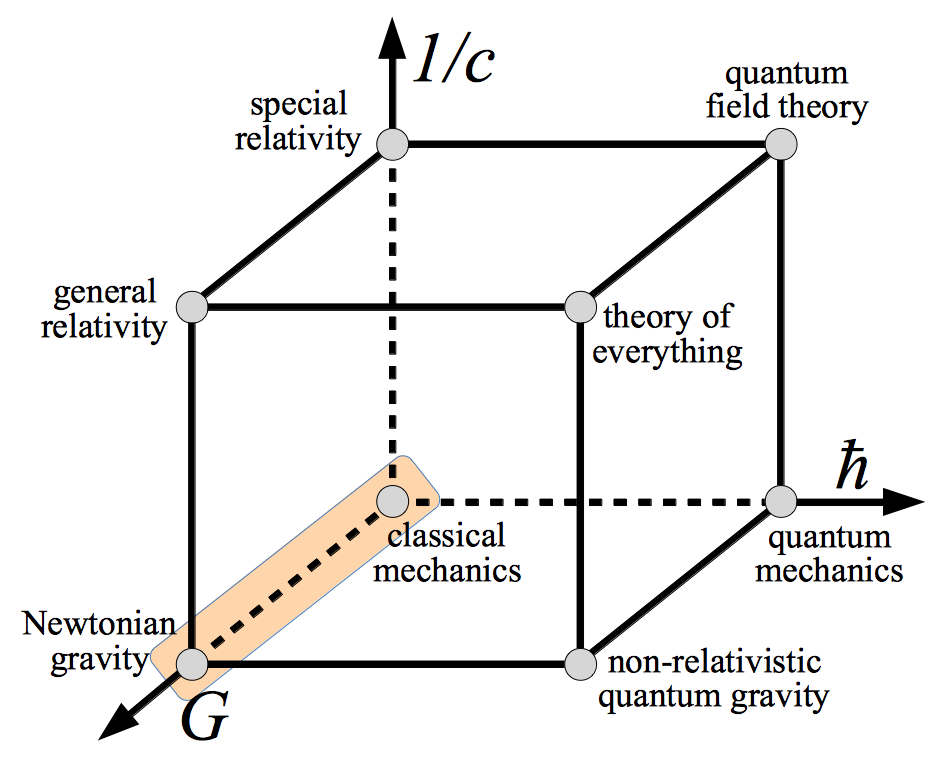}
\caption{\label{fig:cube}{\small Cube of theories \citep{Okun2002}. The origin
corresponds to all physical constants being zero, i.e., $G = \hbar = 1/c = 0$,
while the rest of the vertices represent the different theories in which
gravitational, quantum, relativistic effects, or a combination of them, are
felt and must be accounted for.}}
\end{center}
\end{figure}

In this spirit, I have met a very curious idea and you have probably met it
too: That ``being understood'' is just being intuitive, and that being
intuitive is a feature so much related to everyday experience that \emph{by
definition} we can only ``understand'' classical mechanics. Thus, when I say
that ``I don't understand QM'' I would only be rephrasing the vacuous
complaint that ``QM is not classical mechanics, and that bothers me''. If we
take a look at the ``cube of theories'' (in fig.~\ref{fig:cube}) introduced by
\cite{Gamow2002} as a humorous present to a female student that the three
young friends courted [it seems: \citep{Okun2002}], the argument would state
that understanding is only possible in the small orange region that comprises
classical mechanics with and without gravity. QM is just outside this region
and nothing can be done about it but grow a healthy resignation.

This point of view is very clearly found in some quotes by Feynman:

\begin{quote}
{\small Because atomic behavior is so unlike ordinary experience, it is very
difficult to get used to, and it appears peculiar and mysterious to everyone
---both to the novice and to the experienced physicist. Even the experts do 
not understand it the way that they would like to, and it is perfectly 
reasonable that they should not, because all of direct human experience and 
of human intuition applies to large objects. We know how large objects will 
act, but things on small scale just do not act that way. So we have to learn
about them in a sort of abstract or imaginative fashion and not by connection
with our direct experience. \citep[p.~1-1]{Feynman1963}}
\end{quote}

\begin{quote}
{\small In the beginning of the history of experimental observation on 
scientific things, it is intuition, which is really based on simple experience 
with everyday objects, that suggests reasonable explanations for things. But
as we try to widen and make more consistent our description of what we see, as 
it gets wider and wider and we see a greater range of phenomena, the 
explanations become what we call laws instead of simple explanations. One odd
characteristic is that they often seem to become more and more unreasonable 
and more and more intuitively far from obvious. [\ldots] There is no reason 
why we should expect things to be otherwise, because the things of everyday
experience involve large numbers of particles, or involve things moving very
slowly, or involve other conditions that are special and represent in fact a
limited experience with nature. \citep[p.~127]{Feynman1965}}
\end{quote}

\begin{quote}
{\small Electrons, when they were first discovered, behaved exactly like 
particles or bullets, very simply. Further research showed, from electron 
diffraction experiments for example, that they behaved like waves. As time 
went on there was a growing confusion about how these things really behaved 
---waves or particles, particles or waves? Everything looked like both.

This growing confusion was resolved in 1925 or 1926 with the advent of
the correct equations for quantum mechanics. Now we know how the electrons and
light behave. But what can I call it? If I say they behave like particles I
give the wrong impression; also if I say they behave like waves. They behave
in their own inimitable way, which technically could be called a quantum
mechanical way. They behave in a way that is like nothing that you have seen
before. Your experience with things that you have seen before is incomplete.
The behavior of things on a very tiny scale is simply different. An atom does
not behave like a weight hanging on a spring and oscillating. Nor does it
behave like a miniature representation of the solar system with little planets
going around in orbits. Nor does it appear to be somewhat like a cloud or fog
of some sort surrounding the nucleus. It behaves like nothing you have seen
before. \citep[p.~128]{Feynman1965}}
\end{quote}

But not only him. Van Kampen also seems to think that this is the case:

\begin{quote}
{\small The difficulty is that the authors are unable to adjust their way of 
thinking ---and speaking--- to the fact that phenomena on the microscopic 
scale look different from what we are accustomed to in ordinary life.
\citep{VanKampen2008}}
\end{quote}

In my opinion, the problem with QM is not that it is counterintuitive, it is
not that it is different from classical mechanics. As mentioned in the quotes
above, being counterintuitive is completely expected when the theory tackles
phenomena that are outside everyday experience. QM talks about the very small,
and we never see or experience the very small directly, so it is normal that
we find QM counterintuitive. So far so good.

The problem is that many other theories are counterintuitive, \emph{but they
haven't produced tens of interpretations}. Special relativity is also
counterintuitive, because it deals with bodies that move close to the speed of
light (which is much faster than the fastest of our spacecrafts). There are
some nice counterintuitive effects predicted by special relativity, but there
are hardly more interpretations than one, and the prolific discussions in the
literature with tens of papers being produced each month about how to
understand the quantum mechanical worldview are just \emph{missing} in the
case of special relativity.

I agree with Feynman that there is a psychological resistance to 
counterintuitive theories:

\begin{quote}
{\small The difficulty really is psychological and exists in the perpetual
torment that results from your saying to yourself, ``But how can it be like
that?'' which is a reflection of uncontrolled but utterly vain desire to see
it in terms of something familiar. I \emph{will not} describe it in terms of 
an analogy with something familiar; I will simply describe it.
\citep[p.~129]{Feynman1965}}
\end{quote}

But I think that it is a resistance which can be overcome with time and 
practice. Special relativity is an example of this.

The very different situation that affects the interpretation of the two,
similarly old theories despite the fact that both depart significantly from
classical experience is clearly put by \cite{Kochen2013}:

\begin{quote}
{\small Almost a century after the mathematical formulation of quantum
mechanics, there is still no consensus on the interpretation of the theory.
This may be because quantum mechanics is full of predictions which contradict
our everyday experiences, but then so is another, older theory, namely special
relativity.}
\end{quote}

In fact, Feynman also mentions relativity just after the above paragraph:

\begin{quote}
{\small There was a time when the newspapers said that only twelve men 
understood the theory of relativity. I do not believe there ever was such a 
time. There might have been a time when only one man did, because he was the 
only guy who caught on, before he wrote his paper. But after people read the 
paper a lot of people understood the theory of relativity in some way or 
other, certainly more than twelve. On the other hand, I think I can safely say 
that nobody understands quantum mechanics.}
\end{quote}

OK, Prof. Feynman, then \emph{what is the difference between the two}? To me,
the answer is that the conceptual foundations of QM are simply a mess, while
those of special relativity are not. QM is not yet understood (in the sense of
the word defined in sec.~\ref{subsec:understanding}), but special relativity
is. \emph{This is only possible because understanding has nothing to do with
being intuitive or with being classical.}

The quote by \cite{Streater2007} (of which we already met a part in 
sec.~\ref{subsec:understood}) is a clamorous failure to notice that the 
different level of consensus between the two theories might be indicating 
something relevant.

\begin{quote}
{\small The continuing work on quantum paradoxes parallels a similar output
after relativity was first formulated: when simple paradoxes concerning
time-dilation or Lorentz contraction were explained, more and more elaborate
versions, (twin paradox, thermodynamic paradoxes involving absorption and
emission in a gravitational field\ldots) were claimed to show some
mathematical or physical inconsistency in the special or general theory.
Rebutting these took up the time of the advocates of the theory, and led to
clarification. But the activity died out. The quantum paradox community is,
even after all these years and all the rebuttals, still on the increase. It
has spawned larger enterprises, such as Bohmian mechanics, Nelson dynamics,
geometro-stochastic dynamics, quantum-state reduction, and the quantum
theories of the brain of Penrose and of Stapp. These are lost causes too, and
fall down with the rebuttal of the spooky action at a distance claimed to
inspire them.}
\end{quote}

As we discussed in sec.~\ref{subsec:understood}, Streater prefers to believe
that the reason behind the difference is that the scientists that investigate
QM are specially myopic. ``Time will come when they will all realize that they
are wasting their lives in a lost cause ---but boy is it taking long!''

\subsection{These are the rules of the game}
\label{subsec:rules}

\begin{flushright}
\parbox[h]{5in}{{\small
There are three rules for writing a novel. Unfortunately, no one knows what 
they are.
\begin{flushright}
\vspace{-23.5pt}
--- William Somerset Maugham
\end{flushright}
}}
\end{flushright}

We have already met some examples of ad hoc invented rules and prohibitions
that seem to come from nowhere and whose only aim appears to be protecting
certain ways of thinking, certain statements: ``Understanding is not the job
of physics'', ``Thou shall not philosophize'', ``The puzzle is irreducible'',
etc. In this section, I introduce two more such rules which tend to co-occur
with the ``Shut up and calculate!'' command, and which do not stand for half
a round once you critically scrutinize them.

When I hear an example of such a rule, I often find that a powerful argument
against it is just to answer: ``OK, and why is that?'' But we can also turn
the words by Feynman about philosophers against those physicists that become
part-time lawmakers:

\begin{quote}
{\small \ldots what the characteristics of nature are, are not to be
determined by pompous preconditions, they are determined always by the
material with which we work, by nature herself. We look, and we see what we
find, and we cannot say ahead of time successfully what it is going to look
like. [\ldots] In fact it is necessary for the very existence of science that
minds exist which do not allow that nature must satisfy some preconceived
conditions, like those of our philosopher [physicist?].
\citep[p.~147]{Feynman1965}}
\end{quote}

\subsubsection{The most important thing a theory must do is predict}
\label{subsubsec:predict}

\begin{flushright}
\parbox[h]{5in}{{\small
Prediction is very difficult, especially of the future.
\begin{flushright}
\vspace{-9pt}
--- Niels Bohr (it seems)
\end{flushright}
}}
\end{flushright}

You have probably heard something to the effect of: ``The great power of 
physics and the difference between physics and the rest of sciences is its
predictive power'', ``The most important thing a theory must do is predict'',
``Economics is the science that explains the past'' (derogatively). Let us
read some of our friends about the topic:

\begin{quote}
{\small So we must talk about what we can predict. 
\citep[p.~2-3]{Feynman1963}}
\end{quote}

\begin{quote}
{\small Scientific theories predict the probabilities of outcomes of 
experiments. \citep{Kochen2013}}
\end{quote}

\begin{quote}
{\small The basis of a science is its ability to \emph{predict}. To predict
means to tell what will happen in an experiment that has never been done.
\citep[p.~2-8]{Feynman1963}}
\end{quote}

It amazes me how such incredibly concrete rules can be concocted about a set
of ``things'' so amazingly complex and heterogeneous as scientific theories,
but hey, that doesn't mean that our friends are wrong. Maybe it is true that
prediction is the most precious and absolutely essential property that a
theory must have\ldots or maybe it isn't. We could ask our friends \emph{why}
prediction is the gold standard by which scientific theories must be judged,
but it would be a rhetorical question, because the statement is just
\emph{false}.

Prediction is of course great. It shows that your theory is so powerful, that
the knowledge of the world it provides to you is so accurate that you can
perform the almost magical trick of \emph{knowing what will happen in
advance.} It is said that the ability to predict solar eclipses made native
cultures look at invaders as semi-gods a long time ago. I will not deny that
prediction is a good thing, but ``the basis of a science''? I think that it is
too bold a claim.

For that matter, the conditions that are required for the very act of
prediction are not always present ---and yet, it would seem that in many cases
we are still ``doing science''. Let us read the following paragraph by
Feynman:

\begin{quote}
{\small The problem has been raised: if a tree falls in a forest and there is
nobody there to hear it, does it make a noise? A real tree falling in a real
forest makes a sound, of course, even if nobody is there. Even if no one is
present to hear it, there are other traces left. The sound will shake some
leaves, and if we were careful enough we might find somewhere that some thorn
had rubbed against a leaf and made a tiny scratch that could not be explained
unless we assumed the leaf were vibrating. So in a certain sense we would have
to admit that there is a sound made. \citep[p.~2-8]{Feynman1963}}
\end{quote}

Even if he is not using equations, he is using his knowledge about what sound
is (air pressure waves), how it can move physical objects such as leaves, and
how harder objects than leaves can leave marks on them. By looking at the
marks, he sees fit to conclude that ``we would have to admit that'' a sound
``happened'' in the forest. As the (probably apocryphal) quote by Bohr at the
beginning of this section jokingly puts it, ``prediction'' is typically
understood ``of the future'' ---and yet it seems that Feynman has just used
some informal theory to produce reliable claims \emph{about the past}. The
word he uses is a good name for it: ``explaining''. This is another possible
use that we can give to scientific theories and I don't know why we should say
that this is less important than predicting the future. For example, in the
falling tree case above, not only reproducing the whole ``experiment'' (same
forest, same tree, same thorns, same leaves) would have proved costly, but it
seems that Feynman though it was also \emph{unnecessary}.

If you think that what Feynman did in the quoted paragraph is not important or
if you think it is not science, you only have to learn about his participation
in the investigation about the disaster of the space shuttle ``Challenger'' in
1986 [see for example \citep{Feynman1988}]. In that occasion, Feynman
contributed to \emph{explaining} the reasons of the accident using more
sophisticated science than just a qualitative understanding of sound ---of
course, repeating the experiment was also out of the question in that
occasion. Similarly, you can think about Darwin's evolution; a theory which is
much more valued for its explanatory powers than for its predictions. We could
call all these applications of rigorous rational thinking, all these uses to
which we can put our dearest disciplines such as physics and biology,
``non-scientific'' if we want. In that way, the claim that prediction is the
most important part of science will stand (by definition), but what would be
the point of it?

The truth is that scientific theories have many properties at which we have to
look if we want to decide if we are satisfied with them: predictive power
of course, but also explanatory power, internal consistency, compatibility
with other theories, ability to suggest new experiments, simplicity,
conceptual clarity, etc. We could start an eternal debate about which
properties are necessary, which are the most important ones, and which grant
the use of the distinguished ``scientific'' label, but it would probably be a
waste of time. If you allow a recommendation, whenever you listen that ``QM is
predictive and that's all a proper scientific theory needs to do'', you can
ask ``OK, and why is that?'', and patiently wait for an answer.

\subsubsection{You can only talk about what can be measured}
\label{subsubsec:talk_measured}

\begin{flushright}
\parbox[h]{5in}{{\small
We find no sense in talking about something unless we specify how we measure 
it; a definition by the method of measuring a quantity is the one sure way of 
avoiding talking nonsense.
\begin{flushright}
\vspace{-23.5pt}
--- (Sir) Hermann Bondi
\end{flushright}
}}
\end{flushright}

That only measurable quantities are ``properly defined'', ``make sense'' or
that they are the only things about which is it allowed to talk if we want to
``avoid talking nonsense'' is another of the ``pompous preconditions'' (using
the term by Feynman) that often appear when somebody dares to criticize the
present conceptual state of the theory known as QM.

Let us read Feynman on this:

\begin{quote}
{\small Another thing that people have emphasized since quantum mechanics was
developed is the idea that we should not speak about those things which we
cannot measure. [\ldots] Unless a thing can be defined by measurement, it has
no place in a theory. \citep[p.~2-8]{Feynman1963}}
\end{quote}

So it seems that the proponents of the Copenhagen interpretation are not
content with introducing the concept of a ``measure'' into the rules of QM in
a completely undefined and mystical way ---they also want it to be the only
thing we can talk about!

Successful theories are full of examples of ideal concepts which are not
directly measurable: centers of mass, actions, probability density functions,
Lagrangians, etc. Limiting the concepts about which we can ``talk about'' thus
seems a very bad strategy if we want to make sense of the physical world in
which we live in. Moreover (and again), what is the reason behind the
``pompous precondition''? Is there any reason at all?

I am tired of defeating weak adversaries, so let us read the answer by
Feynman, in the paragraph just after the previous one:

\begin{quote}
{\small It is a careless analysis of the situation. Just because we cannot
\emph{measure} position and momentum precisely does not \emph{a priori} mean
that we \emph{cannot} talk about them. It only means that we \emph{need} not
talk about them. [\ldots] It is not true that we can pursue science completely
by using only those concepts which are directly subject to experiment.
\citep[p.~2-8]{Feynman1963}}
\end{quote}

\subsection{Of moral and epistemic weakness}
\label{subsec:weakness}

An overarching theme which is much related to everything that we have been
discussing is the implication that whoever wants to ``understand'' QM suffers
from a moral or epistemic weakness. As we have seen, it can take many forms:

You can be too demanding, asking for unreasonable things:

\begin{quote}
{\small Actually quantum mechanics provides a complete and adequate 
description of the observed physical phenomena on the atomic scale. What else 
can one wish? \citep{VanKampen2008}}
\end{quote}

You may abandon the righteous path of sound, austere physics lured by the
poetic but utterly useless ``profundities'' and ``low-level baloney'' that are 
common in that dignified hobby known as ``philosophy'':

\begin{quote}
{\small The scandal is that there are still many articles, discussions, and 
textbooks, which advertise various interpretations and philosophical 
profundities. \citep{VanKampen2008}}
\end{quote}

\begin{quote}
{\small In the early fifties I suffered temporarily from a disease of middle 
age: I used to give philosophical talks about science \ldots 
\citep{Feynman1985}}
\end{quote}

\begin{quote}
{\small \ldots and there was all this other stuff which I thought was 
low-level baloney. \citep{Feynman1985}}
\end{quote}

You can mistake a theory having conceptual problems with your own 
psychological inability to accept that a theory is counterintuitive, simply
because you are unable to adjust your thinking and talking:

\begin{quote}
{\small The difficulty is that the authors are unable to adjust their way of 
thinking ---and speaking--- to the fact that phenomena on the microscopic 
scale look different from what we are accustomed to in ordinary life.
\citep{VanKampen2008}}
\end{quote}

\begin{quote}
{\small The difficulty really is psychological and exists in the perpetual
torment that results from your saying to yourself, ``But how can it be like
that?'' which is a reflection of uncontrolled but utterly vain desire to see
it in terms of something familiar. \citep{Feynman1965}}
\end{quote}

You may let your feelings get in the way of your rational thinking:

\begin{quote}
{\small The next question is whether, when trying to guess a new law, we 
should use the seat-of-the-pants feeling and philosophical principles\ldots
\citep{Feynman1965}}
\end{quote}

\begin{quote}
{\small The ``paradox'' is only a conflict between reality and your feeling of 
what reality ``ought to be''. \citep[18-9]{Feynman1963}}
\end{quote}

Or you can be childish and refuse to play by the rules that your wise
scientific parents kindly transmitted to you for the sake of your own safety
and mental well being (see sec.~\ref{subsec:rules}).

If this is so, if these weaknesses plague you, you can repent and begin the
long path to responsibility, self-control and adulthood by reading any solid
textbook which talks about ``measuring'' in the proper way, i.e., treating it
as a completely unproblematic term. Or you can remember that \emph{ad hominem}
attacks and cheap psychological analyses are the signs that someone has run
out of arguments, and just pretend you didn't listen.

\subsection{Do not work in QM's foundations if you want to succeed}
\label{subsec:success}

\begin{flushright}
\parbox[h]{5in}{{\small
The juvenile sea squirt wanders through the sea searching for a suitable rock 
or hunk of coral to cling to and make its home for life. For this task, it has 
a rudimentary nervous system. When it finds its spot and takes root, it 
doesn't need its brain anymore so it eats it! (It's rather like getting 
tenure.)
\begin{flushright}
\vspace{-9pt}
--- \cite{Dennett1991}
\end{flushright}
}}
\end{flushright}

When all persuasion strategies have failed, the only thing that remains to be
tried is to suggest that your career, your assets, your mental well-being,
your ``success'' in general will be jeopardized if you are fool enough to
break the rules and do the forbidden thing. This last-resort (and somewhat
desperate) move is in the air all the time, but it is also explicitly used in
the otherwise great book by \cite{Streater2007}.

The book offers very articulate insights about many topics that have been
mentioned here and it is in my opinion a must to anybody who wants to begin to
make sense of the field, to know what the questions are, and to meet strong
and consistent proposals to answer them. The problem to me is the same as with
the statements by van Kampen discussed in sec.~\ref{subsec:understood}.
Calling ``lost causes'' to open lines of research followed by so many
scientists at the moment seems an instance of what the author himself
criticizes about ``Bohmians'' in p.~107:

\begin{quote}
{\small Bohmians are very wary of jumping to any conclusions; even in easy 
cases.}
\end{quote}

The recommendation about your job future as a scientist ---which is completely
consistent with this attitude--- comes in p.~95:

\begin{quote}
{\small This subject [hidden variable theories of QM] has been thoroughly
worked out and is now understood. A thesis on this topic, even a correct one,
will not get you a job.}
\end{quote}

On the one hand, fear of starvation or being an outcast is no scientific
argument. On the other hand, this is just false. As you can check by scanning
the literature yourself, it is entirely possible to write a thesis in hidden
variables, even an incorrect one, and still get a job ---although certainly
not in Streater's group, it seems. Maybe he is right and hidden variable
theories turn out to be of no value in the end; however, and given that nobody
knows at the moment how to clean up the conceptual QM mess, this path could in
principle be as good as any other. Check for example
\citep{Mermin1993,Schlosshauer2011} for cogent analyses on the present
situation of this line of work, which do not assume from the start that:

\begin{quote}
{\small Bell's theorem, together with the experiments of Aspect et al., shows 
that the theoretical idea to use hidden classical variables to replace quantum 
theory is certainly a lost cause, and has been for forty years.
\citep[p.~99]{Streater2007}}
\end{quote}

About Bohmian mechanics, \cite{Streater2007} has a different message for you:

\begin{quote}
{\small This subject was assessed by the NSF of the USA as follows ``\ldots
The causal interpretation [of Bohm] is inconsistent with experiments which
test Bell's inequalities. Consequently \ldots funding \ldots a research
programme in this area would be unwise''. I agree with this recommendation.}
\end{quote}

So it seems that the situation is slightly better with Bohmian mechanics than
with hidden variables (even if some researchers say that the former is an
example of the latter). Although you will get no funding from the NSF if you
go all Bohmian, at least the possibility that you might get a job is open. On
the other hand, since ``funding agencies are wise'' is a proposition of
uncertain truth value, the above quote is again no scientific argument at all.
As the many interesting works that are published in the field every month
show, doing research in Bohmian mechanics is entirely justified ---that is, if
you can live without NSF funding. If you are interested to begin to explore
this line (even if that means paying for your own conference fees) you can
check the following references: \citep{Durr2009,Nikolic2008,Oriols2012}.
However, be careful, since this would go against the advice in
\citep[p.~112]{Streater2007}:

\begin{quote}
{\small Better steer clear of Bohmians.}
\end{quote}

My advice to you is slightly different: Better steer clear of people that tell
you to steer clear of other people. Although it is a logical contradiction as
stated, I am sure that you can minimally modify it so that it both makes sense
and becomes a hopefully useful advice.

\section{Conclusions}
\label{sec:conclusions}

\begin{flushright}
\parbox[h]{5in}{{\small
A conclusion is the place where you got tired of thinking.
\begin{flushright}
\vspace{-9pt}
--- Martin H. Fischer
\end{flushright}
}}
\end{flushright}

\begin{flushright}
\parbox[h]{5in}{{\small
An adventure is only an inconvenience rightly considered. An inconvenience is 
only an adventure wrongly considered.
\begin{flushright}
\vspace{-23.5pt}
--- G. K. Chesterton
\end{flushright}
}}
\end{flushright}

Since I have completely abused his words, let me quote Feynman again to
begin to wrap up:

\begin{quote}
{\small \ldots mathematicians prepare abstract reasoning ready to be used if
you have a set of axioms about the real world. But the physicist has meaning
to all his phrases. [\ldots] in physics you have to have an understanding of
the connection of words with the real world. It is necessary at the end to
translate what you have figured out into English, into the world, into the
blocks of copper and glass that you are going to do the experiments with.
\citep[p.~55]{Feynman1965}}
\end{quote}

This is completely so in my opinion, and I think that I have presented strong
plausibility arguments supporting that the task of ``understanding the
connection of the words with the real world'' is something which still has to
be achieved in the case of one of humankind's most successful theories:
quantum mechanics.

It is not only very important that we do so, but it is also slightly
embarrassing that we haven't done it yet. In this, I agree with 
\cite{Carroll2013}:

\begin{quote}
{\small Not that we should be spending as much money trying to pinpoint a 
correct understanding of quantum mechanics as we do looking for supersymmetry, 
of course. The appropriate tools are very different. We won't know whether 
supersymmetry is real without performing very costly experiments. For quantum 
mechanics, by contrast, all we really have to do (most people believe) is 
think about it in the right way. No elaborate experiments necessarily required 
(although they could help nudge us in the right direction, no doubt about 
that). But if anything, that makes the embarrassment more acute. All we have 
to do is wrap our brains around the issue, and yet we've failed to do so.}
\end{quote}

But as him\ldots

\begin{quote}
{\small I'm optimistic that we will, however. And I suspect it will take a lot 
fewer than another eighty years. The advance of experimental techniques that 
push the quantum/classical boundary is forcing people to take these issues 
more seriously. I'd like to believe that in the 21st century we'll finally 
develop a convincing and believable understanding of the greatest triumph of 
20th-century physics. \citep{Carroll2013}}
\end{quote}

Now, what we need is clever people that are not afraid of thinking deep, of
getting their hands dirty, and of tackling great problems. If you feel like
giving it a try and you meet some headwind howling that ``everything is
understood'', that ``understanding is not our job'', or that it is
``impossible'' or ``too difficult'', if some colleagues tell you that
``physics does not deal with whys'', that ``understanding is just being
Newtonian'' or that ``Thou shall not philosophize'', if they throw invented
norms to you, or if weaknesses and career prospects enter the discussion when
the debate gets all heated up\ldots you can always answer ``Shut up and let me
think!'' ---and get back to work.

\section*{Acknowledgements}

\hspace{0.5cm} I would like to thank J.~L.~Alonso for illuminating
discussions. Our conversations about quantum mechanics have been long and
many, and I have enjoyed them all ---which of course doesn't mean that our
opinions coincide. I also thank Lucien Hardy for pointing out the very similar
motto introduced by Rob Spekkens and him, Michel Dyakonov and Chris Fields for
noticing that my paper was published the same day as the one by Berthold-Georg
Englert defending just the opposite position, and Giancarlo Ghirardi for
nicely pointing out that I had forgotten to mention collapse models. Finally,
I thank Iv\'an Calvo, Antonio Garc\'{\i}a Cordero, David Zueco and some
anonymous referees at the American Journal of Physics (the first place where I
tried to get this published) for correcting some typos in previous versions of
the manuscript.

This work has been supported by the grants FIS2009-13364-C02-01 (MICINN,
Spain), UZ2012-CIE-06 (Universidad de Zaragoza, Spain), Grupo Consolidado
``Biocomputaci\'on y F\'{\i}sica de Sistemas Complejos'' (DGA, Spain).


\end{document}